# Negative group velocity in layer-by-layer chiral photonic crystals

Kin Hung Fung, Jeffrey Chi Wai Lee, and C. T. Chan

Department of Physics and William Mong Institute of Nano Science and Technology,

The Hong Kong University of Science and Technology, Hong Kong, China

We study the group velocity of light in layer-by-layer chiral photonic crystals composed of dielectrics and metals. Through studying the band structures with an extended-zone scheme that is given by a Fourier analysis, we show the existence of negative group velocity in the proposed chiral structures. The physical mechanism is interpreted with the help of a simplified model that has an analytical solution. The iso-frequency contours of the photonic band structure suggest that the negative group velocity can lead to either positive or negative refraction, depending on the orientation of the medium interface. We propose a feasible realization of such kind of photonic crystals. Computational results on the proposed realization are consistent with that of the simplified models.

## I. INTRODUCTION

Artificial chiral photonic structures have drawn increasing interests in recent years because they can support special photonic properties, such as polarization gap (or circular Bragg gap),[1,2,3,4] high fidelity defect mode,[5,6,7] negative group velocity and negative refraction.[8 9 10 11 12 13 14] Negative group velocity is of special interest as it is closely related to negative refraction and near-field super-resolution focusing.[15] One way to achieve negative group velocity is to employ Veselago's idea of a double negative medium,[16] which has simultaneous negative electric permittivity and negative magnetic permeability (i.e., double negativity).

Double negativity can be experimentally realized in metamaterials that are composed of electric and magnetic resonators. The first realization of double negative metamaterials by Shelby *et al.*[17] has stimulated rapid growth in the research on resonant materials. Recently, it has been proposed that negative group velocity for one circular polarization may also exist in optically active (chiral) materials that possess electric resonance only.[8] The electromagnetic fields of such kind of materials satisfy the following constitutive relations:[18]

$$\begin{cases} \mathbf{D} = \varepsilon_0 \varepsilon_{eff} \mathbf{E} - i\kappa_{eff} \mathbf{H}/c \\ \mathbf{B} = i\kappa_{eff} \mathbf{E}/c + \mu_0 \mu_{eff} \mathbf{H} \end{cases} \quad (1)$$

where $\varepsilon_0$ is the vacuum electric permittivity, $\mu_0$ is the vacuum magnetic permeability, $c = 1/\sqrt{\varepsilon_0 \mu_0}$, $\varepsilon_{eff}$ is the (effective) dielectric function, $\mu_{eff}$ is the (effective) relative permeability, and $\kappa_{eff}$ is the (effective) coefficient for describing the optical activity. If the medium is homogeneous, wave can propagate with a negative group velocity (i.e., the group velocity and the phase velocity pointing in opposite directions) at the frequency range when [9 14 19 20]

$$|\kappa_{eff}| > \sqrt{\varepsilon_{eff}} \sqrt{\mu_{eff}} \,. \quad (2)$$

To achieve this, Pendry proposed a metamaterial consisting of "Swiss roll" resonators.[8] Such metamaterial possesses a non-zero effective $\kappa_{eff}$ and a Lorentzian resonance in the effective $\varepsilon_{eff}$ so that Eq. (2) can be satisfied at the frequencies where $\varepsilon_{eff}(\omega)$ is close to zero. However, the proposed metamaterial is rather difficult to make because of the small size of the chiral resonators. In spite of this, it has been shown that similar effect can also be achieved in arrays of non-chiral resonators forming a chiral lattice, in which effective medium description may not be valid.[13,21] For convenience, we call such kind of periodic chiral inhomogeneous medium the chiral photonic crystal (PC). If a chiral PC contains resonators, we call it a resonant chiral PC. To facilitate the understanding of the effect of chirality on resonance, we study a simplified resonant chiral PC analytically. We then transform the analytical model to some more realizable layer-by-layer structures and verify that the salient features remain the same using numerical calculations.

This paper is organized as follows. We first study a resonant chiral PC that is formed by continuous twisting of anisotropic medium (Section II). Such PC possesses analytic solution so that band structures in an extended zone scheme can be easily calculated. A Fourier analysis allows us to distinguish the "artificial" negative group velocity (that is due to band folding) from the real negative group velocity (that is due to the chiral effect). Next, we introduce a chiral PC that is similar to the first one except it consists of discrete layers of anisotropic media. We demonstrate the ability of separating lights of different circular polarizations by the layer-by-layer chiral structure (Sec. III). Then, we study the effect of resonances in such structure (Sec. IV). The band structures are compared with those of the continuous one. The relation between negative refraction and negative group velocity is also studied (Sec. V). Finally, we propose a realization of a resonant chiral PC consisting of dielectric bars and metal plates (Sec. VI). Computational results for the proposed structures are compared with the simplified models.

## II. EFFECT OF ELECTRIC RESONANCE ON CHIRAL STRUCTURE

For a direct comparison with the results in this paper, a typical band structure showing Pendry's homogenous chiral medium is reproduced in Figure 1. For the sake of simplicity, we consider only non-magnetic metamaterials ($\mu_{eff} = 1$). The effective dielectric function of the material is

$$\varepsilon_{eff} = 1 + s\omega_0^2(\omega_0^2 - \omega^2)^{-1}, \tag{3}$$

where $\omega_0$ is the resonant frequency and $s$ is the strength of the resonance. By comparing Fig. 1(a) for $\kappa_{eff} = 0$ and Fig. 1(b) for $\kappa_{eff} \neq 0$, we can see the splitting of bands and the appearance of a band with a negative group velocity (opposite to the phase velocity) at the first pass band above $\omega_0$, in which Eq. (2) is satisfied. If the metamaterial is isotropic, negative refraction at the medium interface can be realized. However, the proposed metamaterial would be difficult to make (as discussed in Sec. I). Here, instead of searching for effective medium of non-zero $\kappa_{eff}$, we consider chiral PCs in which the functionality does not require an effective medium description.

We begin by considering a continuously twisted anisotropic medium (similar to cholesteric liquid crystal[1]) with the dielectric tensor of the form,

$$\varepsilon(z) = \mathbf{R}\left(\frac{\pi z}{a}\right) \cdot \varepsilon_0 \cdot \mathbf{R}^{-1}\left(\frac{\pi z}{a}\right), \tag{4}$$

where

$$\mathbf{R}(\theta) = \begin{pmatrix} \cos\theta & -\sin\theta & 0 \\ \sin\theta & \cos\theta & 0 \\ 0 & 0 & 1 \end{pmatrix}, \tag{5}$$

$$\varepsilon_0 = \begin{pmatrix} \varepsilon_1 & 0 & 0 \\ 0 & \varepsilon_2 & 0 \\ 0 & 0 & \varepsilon_3 \end{pmatrix}, \tag{6}$$

in Cartesian coordinates, $\theta$ is the orientation angle in the x-y plane, $\varepsilon_1(\omega)$, $\varepsilon_2(\omega)$, and $\varepsilon_3(\omega)$ are frequency dependent dielectric functions corresponding to the principal axes. The closed-form dispersion relation is[22]

$$(k'-1+2m)^2 = 1 + 2(\varepsilon_1 + \varepsilon_2)\omega'^2 \pm 2\omega'\sqrt{2(\varepsilon_1 + \varepsilon_2) + (\varepsilon_1 - \varepsilon_2)^2 \omega'^2}, \tag{7}$$

where $k' \equiv ka/\pi$ and $\omega' \equiv \omega a/2\pi c \equiv fa/c$ are, respectively, the normalized Bloch wavenumber and the normalized frequency. Here, we have an arbitrary integer, $m$, for choosing the Brillouin zone. The chiral PC considered here has a right-hand structure that breaks the chiral symmetry.

To introduce resonance in the model, one simple way is to use a Lorentzian dielectric function. For example, we take $\varepsilon_1 = \varepsilon_3 = 1$, $\varepsilon_2 = 1 + s\omega_0^2(\omega_0^2 - \omega^2)^{-1}$, and $\omega_0' \equiv \omega_0 a/2\pi c = 0.3$. The band structure in reduced zone scheme is shown in Fig. 2(b). For better comparison, we also plot Fig. 2(a), which is the same as the isotropic non-chiral case shown in Fig. 1(a) but using a reduced zone scheme that is the same as Fig. 2(b). We see that there exists a chirality-induced negative-slope band (located near $\omega \approx 1.47\omega_0$ when $s = 2.78$) above the resonant bandgap. Such phenomenon is very similar to that in the effective medium considered by Pendry in Ref. [8].

The reduced zone scheme is commonly used for showing the band structures of periodic crystals. However, as we will see in the following, the band structure in the reduced zone scheme does not give all the information on the "true" value of $k$, which determines the relative sign between the phase velocity and the group velocity. To facilitate the understanding of the resonant chiral PC, we calculate the band structure in an extended zone scheme. In Eq. (7), $(k'-1+2m)^2$ has two functional forms (in terms of $\omega'$) and, therefore, there are four dispersion formulae for each fixed $m$ [after taking a square root on both sides of Eq. (7)]. We choose a particular $m$ for each formula such that each dispersion relation satisfies the quasistatic limit, $\lim_{\omega' \to 0} k'(\omega') = 0$. Results for some particular resonance strengths, $s$, are shown in Figs. 3(a) and 3(b). The polarization of the electric field for each pass band is also indicated in the figure. Although the polarization of each band is not exactly circular but are generally elliptical, we still call the elliptically polarized light the right-hand (left-hand) polarized light for the sake of simplicity if their electric fields rotate in the clockwise (anti-clockwise) direction when the group (energy) velocity is pointing towards the observer's eye. Such definition is consistent with the conventional definition of polarizations for light propagating in free space. For showing the actual polarization of each mode clearly, we find the time-varying elliptical trajectory of the electric field on the $x-y$ plane and calculate the ratio $E_{min} / E_{maj}$, where $E_{min}$ and $E_{maj}$ are the minor and major axes of the ellipse [see Figs. 3(c) and 3(d)].

At low frequencies ($\omega \ll \omega_0$), the dispersions in Figs. 3(a) and 3(b) are linear and can be approximated as $\omega = (1+s/2)^{-1/2} ck$ ($< ck$) for both polarizations. At frequency close to $\omega_0$, there are hybridizations among local resonance ($\omega = \omega_0$) and propagating modes for both polarizations, opening a resonant band gap just above the resonance frequency $\omega_0$. Apart from the resonant gap, the dispersion curves for the two polarizations show very different behaviors. Since we have a right-hand chiral structure, the right-hand wave can "see" the structure so that the group velocity is zero at the zone boundaries, similar to typical Bragg gaps. In contrast, the left-hand wave show continuous increase in frequency (with non-zero group velocity) as $k$ increases and passes through zone boundaries, except in a region just above the resonant gap, which will be discussed as follows. We find that both Figs. 3(a) and 3(b) show crossing of bands at the zone center at a frequency $\omega_U > \omega_0$, where $\omega_U$ is given by

$$\omega_U'^2 = \frac{1}{2}\left[1 + \omega_0'^2 + s\omega_0'^2 - \sqrt{(1-\omega_0'^2)^2 + (2+s)s\omega_0'^4}\right], \tag{8}$$

and $\omega_U' = \omega_U a / 2\pi c$. For $s=4$ [Fig. 3(a)] and $s=12$ [Fig. 3(b)], we have $\omega_U \approx 1.56\omega_0$ and $1.96\omega_0$, respectively. Below such band-crossing point, the figure clearly shows a left-hand polarization band with negative slope. For a weaker resonant strength [see Fig. 3(a)], such negative-slope band is connected with a positive-slope band of the right-hand polarization at a polarization cross-over point of the frequency, $\omega_L$, which is given by

$$\omega_L'^2 = \omega_0'^2\left[1 + \frac{s}{8}\left(2 - s\omega_0'^2 + \sqrt{(2-s\omega_0'^2)^2 - 16\omega_0'^2}\right)\right], \tag{9}$$

where $\omega'_L \equiv \omega_L a / 2\pi c$. For $s = 4$, we have $\omega_L \approx 1.54\omega_0$. (It should be noted that the rotational direction of the electric field in the $x-y$ plane is preserved across the transition point. The change in polarization is due to the change in the propagation direction of the wave, i.e. the group velocity.)

The bandwidth of the negative-slope band is, therefore, $\Delta\omega = \omega_U - \omega_L$, which is about $0.02\omega_0$ for $s = 4$. For a larger resonant strength, $\Delta\omega$ can be increased. If $s > s_C \equiv 2(1 - 2\omega'_0)/\omega'^2_0$, $\omega_L$ will become complex and the whole band will belong to one single (left-hand) polarization only. In this case, the lower band-edge frequency is $\omega_B = \pi c / a$, instead of $\omega_L$. In our examples, we have $s_C \approx 8.89$. Therefore, no right-hand polarization band appears near the negative-slope band in Fig. 3(b), and the bandwidth of the negative-slope band becomes $\Delta\omega = \omega_U - \omega_B$. For $s = 12$, we have $\omega_B \approx 1.67\omega_0$ and $\Delta\omega \approx 0.29\omega_0$.

Although the band structure in such extended zone scheme satisfies the quasistatic limit, $\lim_{\omega' \to 0} k'(\omega') = 0$, we still cannot conclude from the figure that the band with negative slope corresponds to a negative group velocity. For a periodic inhomogeneous system, the group velocity ($d\omega / dk$) is well defined while the phase velocity ($\omega / k$) is not. It is because the freedom in the choice of the most suitable Brillouin zone (i.e., the exact value of $k$) may lead to an uncertainty in the phase velocity. In this case, a Fourier analysis proposed by Lombardet et al.[23] can be used to tackle such problem. For each photonic mode, we make a Fourier transform of the corresponding electric field and find the corresponding wavenumbers and integrated strengths of the Fourier peaks. Fig. 4 shows the graphs of the frequency versus the wavenumber of the Fourier peaks. The intensity and thickness of the curves in Fig. 4 indicates the integrated strengths of the Fourier peaks. Although there are many values of $k$ (up to multiples of reciprocal lattice vectors) for a given mode, the phase velocity of the mode can still be well approximated using the dominant value of $k$ (i.e. the corresponding value of $k$ of the mode with highest Fourier strength). For the crossing bands near $\omega_U$, the zone-center Fourier component is the dominant component. Therefore, the existence of the negative band below $\omega_U$ is not an artificial band-folding effect[24] but a *bona-fide* chirally induced band of negative group velocity because the sign of the dominant $k$ (i.e. the sign of the phase velocity) is opposite to the group velocity. Fig. 4(a) shows that, for a smaller resonant strength, the Fourier component is more concentrated at zone center but with narrower negative band. In contrast, for a large resonant strength [see Fig. 4(b)], the Fourier strength in the 1st Brillouin zone is only slightly larger than that in the 2nd Brillouin zone.

## III. A SIMPLE CHIRAL STRUCTURE: LAYER-BY-LAYER PHOTONIC CRYSTAL

We have shown in Sec. II that a simple model of continuously twisted anisotropic material can give a negative group velocity when resonance is coupled with chirality. The continuous model can help us understand the phenomena but it is not easy to make because of its perfect continuity. The first step to simplify the chiral structure is to make the structural rotation along the $z$-axis discrete. Here, we will consider a chiral photonic crystal constructed by stacking layers of anisotropic materials. In this example, we stack the anisotropic layers periodically as in Fig. 5. Such photonic crystal has three layers in each period. The first, second and third layer have the dielectric tensors, $\varepsilon(0)$, $\varepsilon(\pi/3)$ and $\varepsilon(2\pi/3)$, respectively. A top view [see Fig. 5(a)] of the structure shows an anti-clockwise 60° rotation for each new layer if we stack the layers from bottom to top. Similar to its continuous counterpart, this right-hand structure has strong rotary power[25] and possesses polarization gaps[1,2,3,4] (or circular Bragg gaps) when both $\varepsilon_1$ and $\varepsilon_2$ are positive. Fig. 6 shows the band structure and the transmission spectra for both right-hand and left-hand polarized light when $\varepsilon_1 = 1$ and $\varepsilon_2 = 1.6$. The results are calculated using the Berreman 4×4 transfer matrix method[26,27] for anisotropic media. Each pass band splits into two, and the splitting is very conspicuous near the band gap. One of the band opens a gap at a frequency $f \approx 0.45c/a$ while the other band opens a gap at nearly twice the frequency $f \approx 0.9c/a$. The transmission spectra for different circular polarizations indicate that each band corresponds to one circular polarization. The results show that such simple model already possesses chiral phenomena, such as the ability of blocking one kind of circularly polarized lights (by the polarization bandgap) and to rotate the polarization angle of linearly polarized light (by the difference in the phase velocities of light of different circular polarizations). One realization of such layer-by-layer chiral PC has been shown to have thermal radiation of circularly polarized light due to its special polarization gap.[28] Since the band structure and transmission properties of such realization can be well explained by such a simple model, we will use this simple chiral structure to study its coupling to material electric resonances in the next section.

# IV. CHIRAL LAYER-BY-LAYER PHOTONIC CRYSTAL WITH MATERIAL ELECTRIC RESONANCE ($\varepsilon$ RESONANCE)

Here, we add a material electric resonance (same as the continuously twisted example in Sec. II) to the 60° layer-by-layer photonic crystal proposed in Sec. III. Fig. 7 shows a comparison between the band structures of these two kinds of photonic crystals for $s = 2.78$. The two band structures are almost the same to the eye, except a small blueshift in frequency. If we enlarge the first band above the resonant bandgap [shown in Fig. 8(a)], we can see a band of negative group velocity appears at $\omega \approx 1.47\omega_0$. The results for larger resonance strengths, $s = 4$, $s = 8$ and $s = 12$ are plotted in Fig. 8(b), 8(c) and 8(d), respectively. We see that, as $s$ increases, the first band above resonance gap shifts to higher frequencies, and the minimum point of the band shifts towards the zone boundary. The result that a stronger resonance gives a wider band of negative slope is the same as the case shown in Fig. 3. The similarity between the continuously twisted structure and such discrete layer-by-layer structure suggests that we can choose either one of them to realize the idea of negative group velocity. In view of this, we propose a layer-by-layer three-dimensional structure for an explicit realization (details are given in Sec. VI).

# V. THE SIGN OF REFRACTION

Before we consider the real structure in Sec. VI, let us use the iso-frequency contour analysis[29] to explain the relationship between negative slope in band structure and negative refraction. Such analysis can be used to determine the direction of the refraction between two media, and its basic principles are briefly outlined as follows. For an incident light beam (with wavevector $\mathbf{k}^{(1)}$) coming from the first medium, we match the component of the wavevector parallel to the interface of the first medium, $\mathbf{k}_\parallel^{(1)}$, to that of the second medium, $\mathbf{k}_\parallel^{(2)}$. Then, we determine the vector group velocities, $\mathbf{v}_g^{(2)} = d\omega/d\mathbf{k}^{(2)}$, of the $\mathbf{k}_\parallel$-matched modes in the second medium by comparing the iso-frequency contour that has a slightly increased frequency. There can be more than one $\mathbf{k}_\parallel$-matched modes for each given $\mathbf{k}_\parallel$. To satisfy the causality, the actual refracted beam should be parallel to the group velocity, $\mathbf{v}_g^{(2)}$, that points towards the second medium [i.e., $(\mathbf{v}_g^{(1)} \cdot \hat{\mathbf{n}})(\mathbf{v}_g^{(2)} \cdot \hat{\mathbf{n}}) > 0$, where $\mathbf{v}_g^{(1)}$ is the group velocity of the incident wave and $\hat{\mathbf{n}}$ is the unit vector normal to the medium interface].

In general, we should plot the iso-frequency contours in the extended zone scheme (as in Sec. II). For the bands just above the resonant frequency, it is enough to focus on the first Brillouin zone because the dominant Fourier peaks are located at the first Brillouin zone (as shown Fig. 4). Fig. 9 shows the iso-frequency contours of the dispersion surface of the 60° layer-by-layer PC for $s = 8$ and $s = 12$ at some frequencies within the band of negative group velocity. (The contours on the $k_y - k_z$ plane are skipped because they are very similar to those on the $k_x - k_z$ plane.) For the weaker resonance case [see Fig. 9(a)], the contours consist of two separate loops, making a complicated refraction of wave. For example, there are four $\mathbf{k}_\parallel$-matched modes when we consider small angle refraction from air to the chiral PC with a medium interface parallel to the $x-y$ plane, but only two of them are casual solutions. These two solutions are of different polarizations [as indicated by the colors of the arrows in Fig. 9(a)]. Therefore, the refraction angle depends on the polarization of the incident wave, but the sign of refraction is positive for both polarizations. For an interface parallel to the $x-z$ plane, both positive and negative refraction can take place, depending on the value of $\mathbf{k}_{\parallel}^{(1)}$ and provided that the $\mathbf{k}_{\parallel}^{(1)}$ falls in the window where there exists $\mathbf{k}_{\parallel}^{(2)}$ such that $\mathbf{k}_{\parallel}^{(2)} = \mathbf{k}_{\parallel}^{(1)}$. The refraction becomes simpler when we have a stronger resonance. In Fig. 9(b), we have only two $\mathbf{k}_{\parallel}$-matched modes for an interface parallel to the $x-y$ plane. Therefore, only one polarization can refract into the chiral PC and the sign of refraction is positive. Similar to the case for weaker resonance, there can be negative refraction for an interface parallel to the $x-z$ plane.

## VI. REALIZATION

In the previous sections, we have studied the effect of introducing electric resonances to a chiral PC using simple models. Here, we realize the idea by a realistic chiral PC as shown in Fig. 10. A unit cell of such chiral PC consists of 3 sub-cells and each sub-cell is made by a layer of parallel and equally spaced dielectric rods (with a period of $a$, a height of $d = 5a/18$, a width of $b = 2a/5$) placed on top of a metal plate (with thickness $e = a/18$). The second sub-cell is placed on top of the first one with a 60° rotation and the third sub-cell is 60° rotated further. The PC repeats itself every 3 layers. Therefore, the period in the $z-$direction is also $a$. The dielectric constant of the rods is 8 while the dielectric function of the metal is given by the lossless Drude model, $\varepsilon(\omega) = 1 - \omega_p^2/\omega^2$, with $\omega_p a/c = 21$, where $\omega_p$ is the plasmon frequency. Both the rods and the metal plates are non-absorbing for the study of the band-structure. The band structure in Fig. 11, calculated by the Plane Wave based Scattering Matrix Method,[30,31] shows pass bands ($\omega \approx 3.22 c/a$) that are very similar to those crossing bands predicted in the previous sections. Furthermore, the iso-frequency contours (as shown in Fig. 12) at $\omega \approx 3.22 c/a$ agree well that in Fig. 9(a). There is thus a close connection of this real chiral PC to those simplified analytical models in previous sections.

Although the results shown in this section show many similarities to those in the previous sections, it should be noted that the physical origin of the negative group velocity is slightly different. In Sec. IV, the negative group velocity is a result of chiral stacking of anisotropic layers that has electric resonant effect included in the dielectric tensor. For the rod-on-plate chiral PC considered in this section, the anisotropy comes from the layers of dielectric rods while the electric resonant response comes from the metal plates.[32] It should also be noted that the effects in this resonant chiral PC are different from those effects in simple non-chiral metal-dielectric stacking.[33] The negative refraction (without negative group velocity) in the later comes from the coupled surface plasmons on the metal plates. In our chiral PC, we focus on the frequency regime in which there is minimal coupling of surface plasmons on the metal plates. In this case, negative group velocity, which can give both positive and negative refractions, comes from the chirality of the structure.

## VII. CONCLUSION

We have studied some alternative ways to add electric resonant effect to chiral photonic structures so that negative group velocity can be achieved. The band structures of an analytical model of resonant chiral PC are studied. Using a Fourier analysis, we have shown that there exists a range of frequency where the group velocity of one elliptical polarization is "truly negative" (i.e., not fictitious negative group velocity due to band folding). We have simplified the analytical model to a discrete layer-by-layer structure and we finally proposed a layer-by-layer metallic-dielectric structure that can be realized experimentally. We have also analyzed the refraction properties through these chiral structures. Results show that the negative group velocity in these structures can lead to both positive and negative refractions. The numerical results for the layer-by-layer structures show many similarities to the analytical model.

In this paper, the proposed effect of negative group velocity is studied beyond the effective medium descriptions in the literature and the results are independent of whether they can be interpreted using effective medium parameters.

## ACKNOWLEDGMENTS

This work was supported by the Central Allocation Grant from the Hong Kong RGC through HKUST3/06C. Computation resources were supported by the Shun Hing Education and Charity Fund.

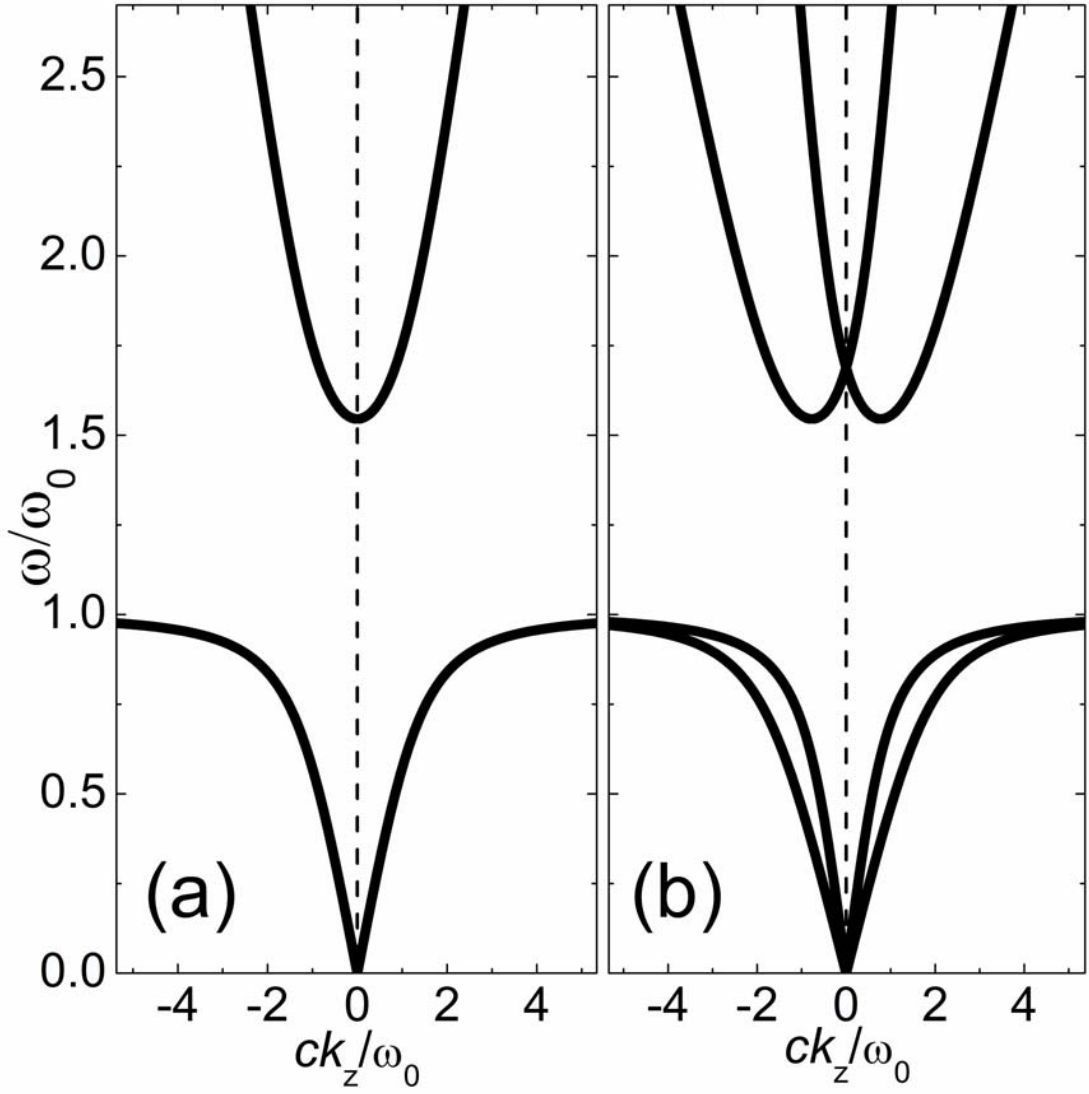

**FIG. 1: Band structures of homogenous chiral and non-chiral media with built-in electric resonance. (a) Homogeneous non-chiral ($\kappa_{eff}=0$) medium, $s=1.39$. (b) Homogeneous chiral ($\kappa_{eff}>0$) medium, $s=1.39$.**

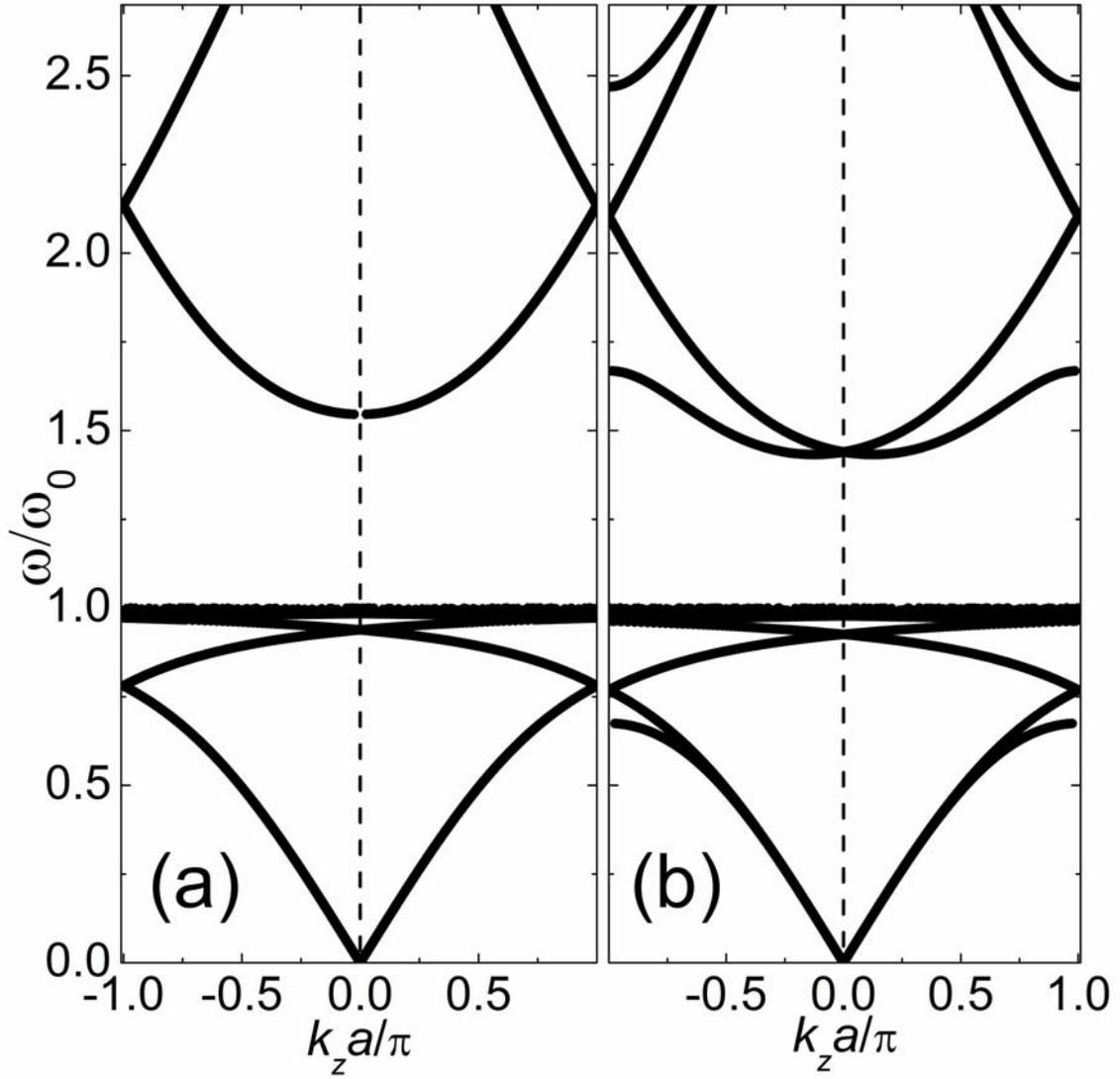

**FIG. 2: Band structures of resonant non-chiral and chiral PCs. (a) Same as Fig. 1(a) except that the dispersions are displayed using the reduced zone scheme. (b) The proposed example of continuously twisted chiral PC with electric resonance ($s = 2.78$).**

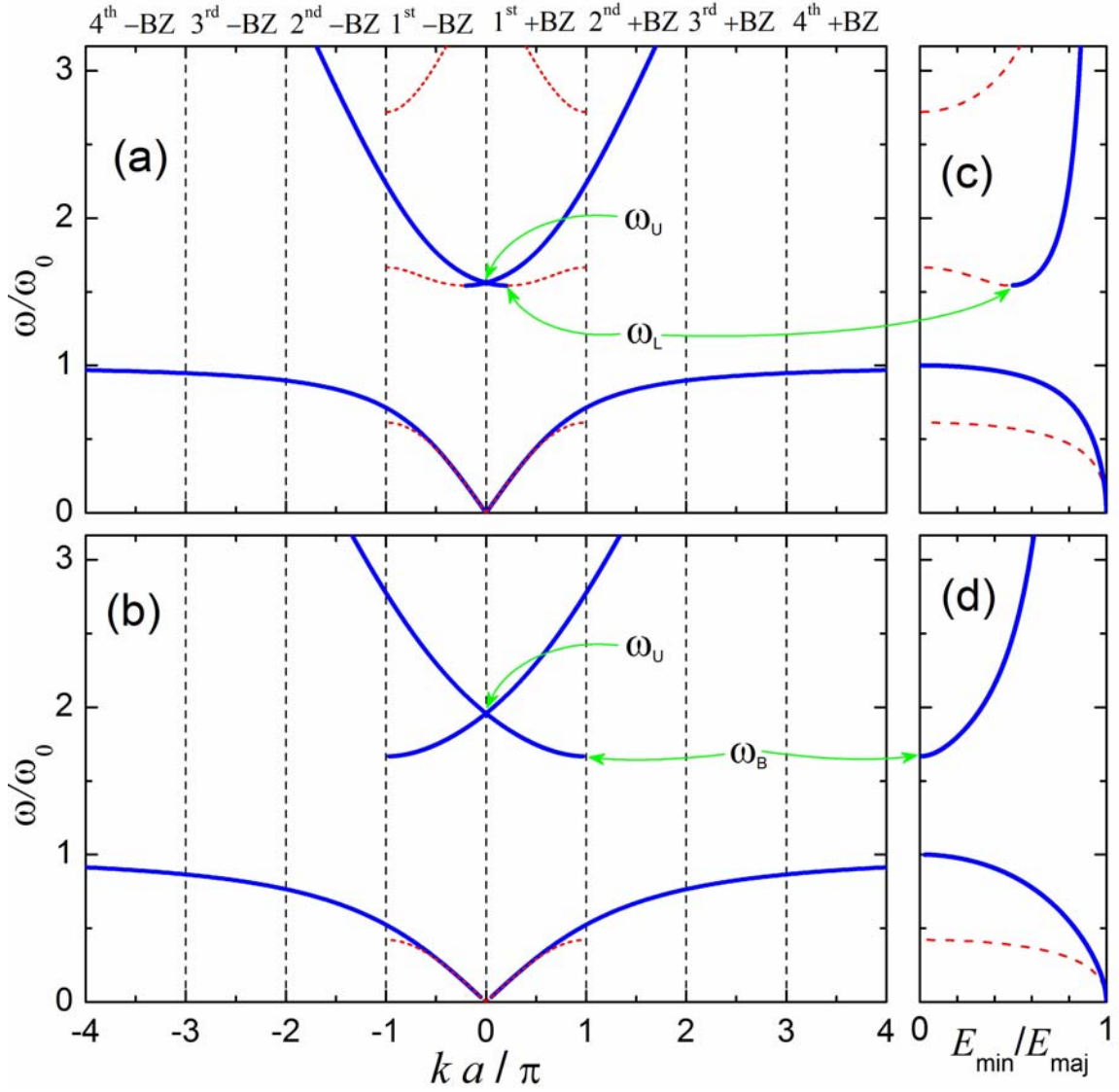

**FIG. 3: (Color online) Band structure of the proposed continuously twisted artificial resonant chiral PC (calculated using the analytic approach). (a) $s=4$ (b) $s=12$. Panels (c) and (d) show the corresponding polarizations of the electric fields for $s=4$ and $s=12$, respectively. Blue solid (red dashed) lines correspond to the left(right)-hand polarization.**

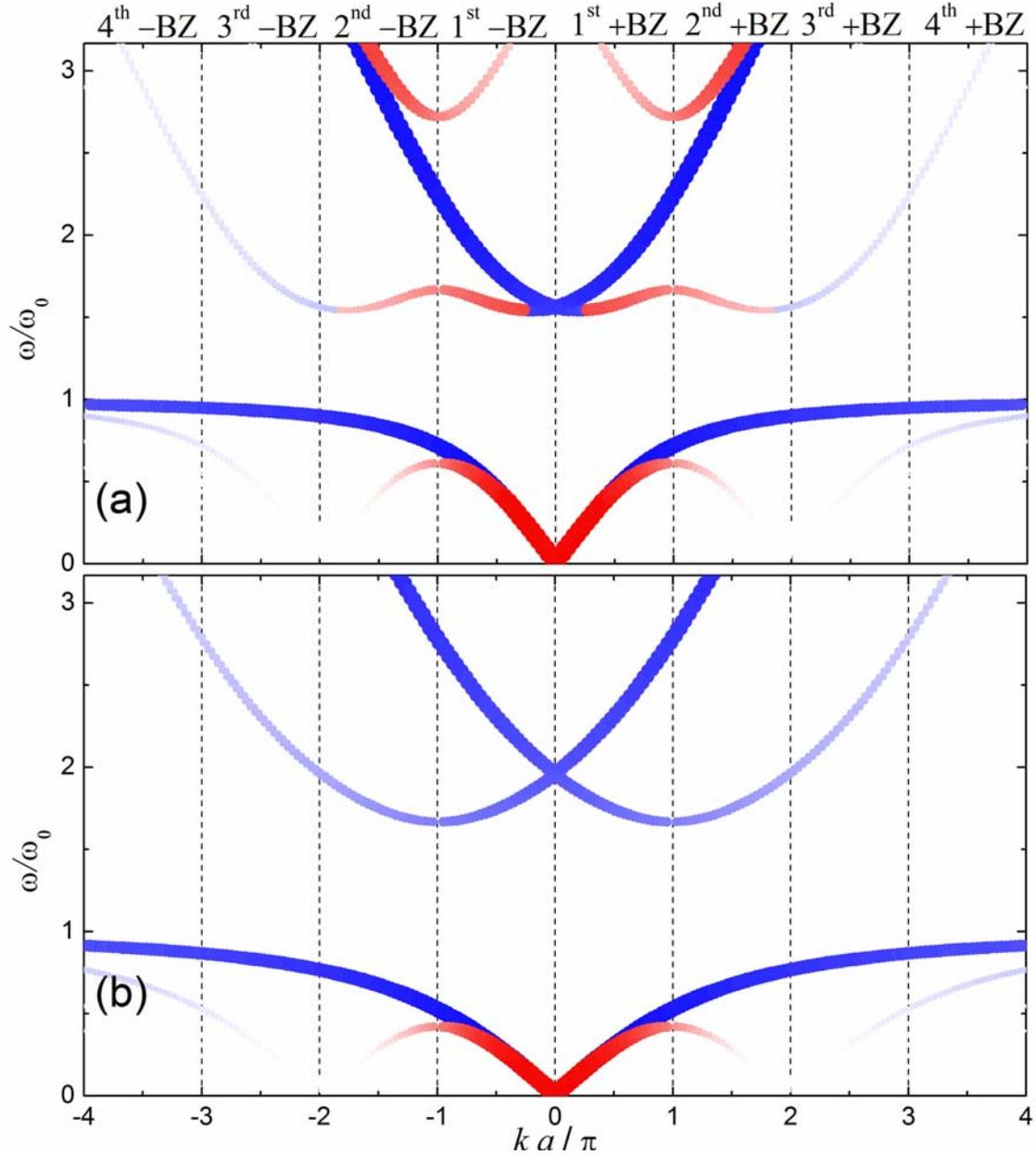

FIG. 4: (Color online) Fourier-intensity plot of the band structure in Fig. 3 in the extended zone scheme. The darkness and the thickness of pass bands show the strength of the Fourier component. (a) $s = 4$ (b) $s = 12$. Blue (red) lines correspond to the left(right)-hand polarization

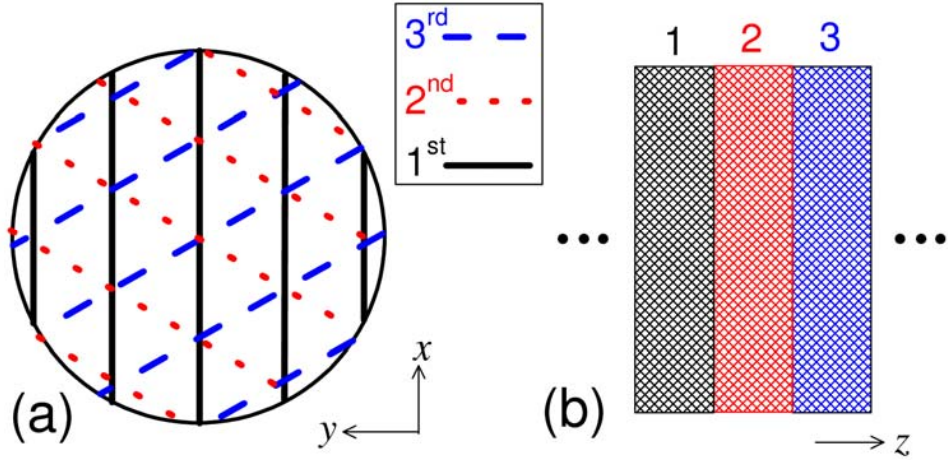

**FIG. 5: (Color online) Schematic structure of the repeating unit of the simplest artificial chiral PC. (a) Top view. The solid (black), dotted (red) and dashed (blue) lines represent, respectively, the first, second and third anisotropic layers in each repeating unit. The lines show also the direction of the anisotropy (one of the principle axis). (b) Side view.**

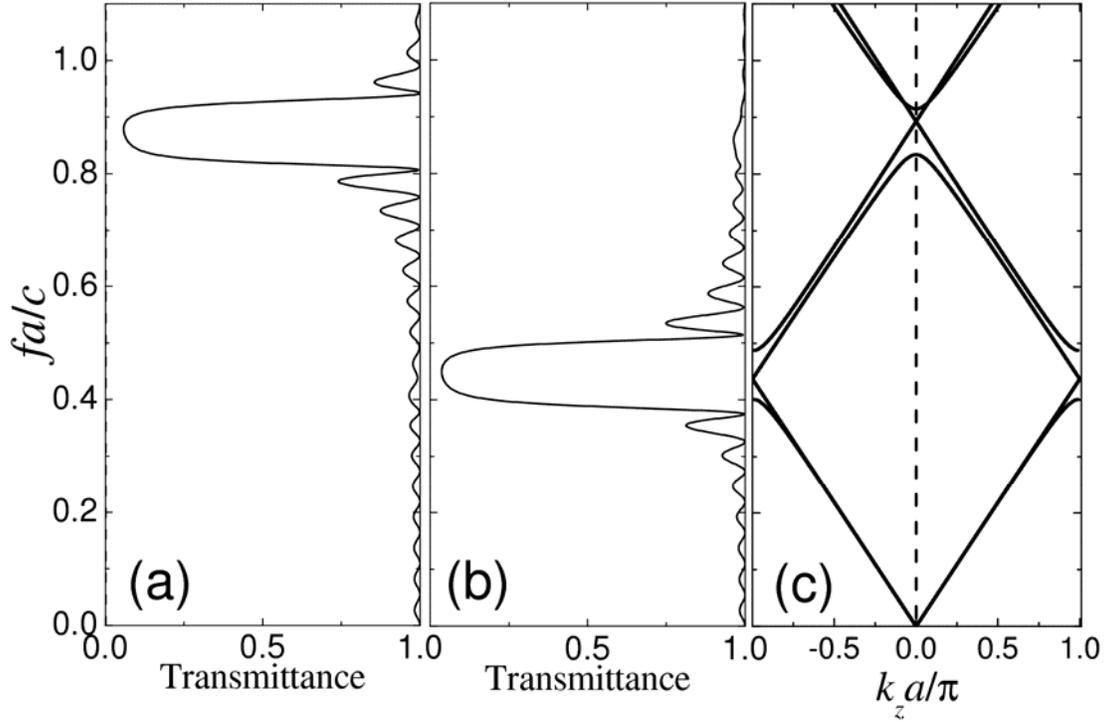

FIG. 6: Properties of the layer-by-layer artificial chiral PC as shown in Fig. 5. (a) Transmission spectrum through 8 periods (24 layers) for left-hand polarized light at normal incidence. (b) The same as (a) but for right-hand polarized light. (c) Band structure along the $z$-direction. $f \equiv \omega/2\pi$.

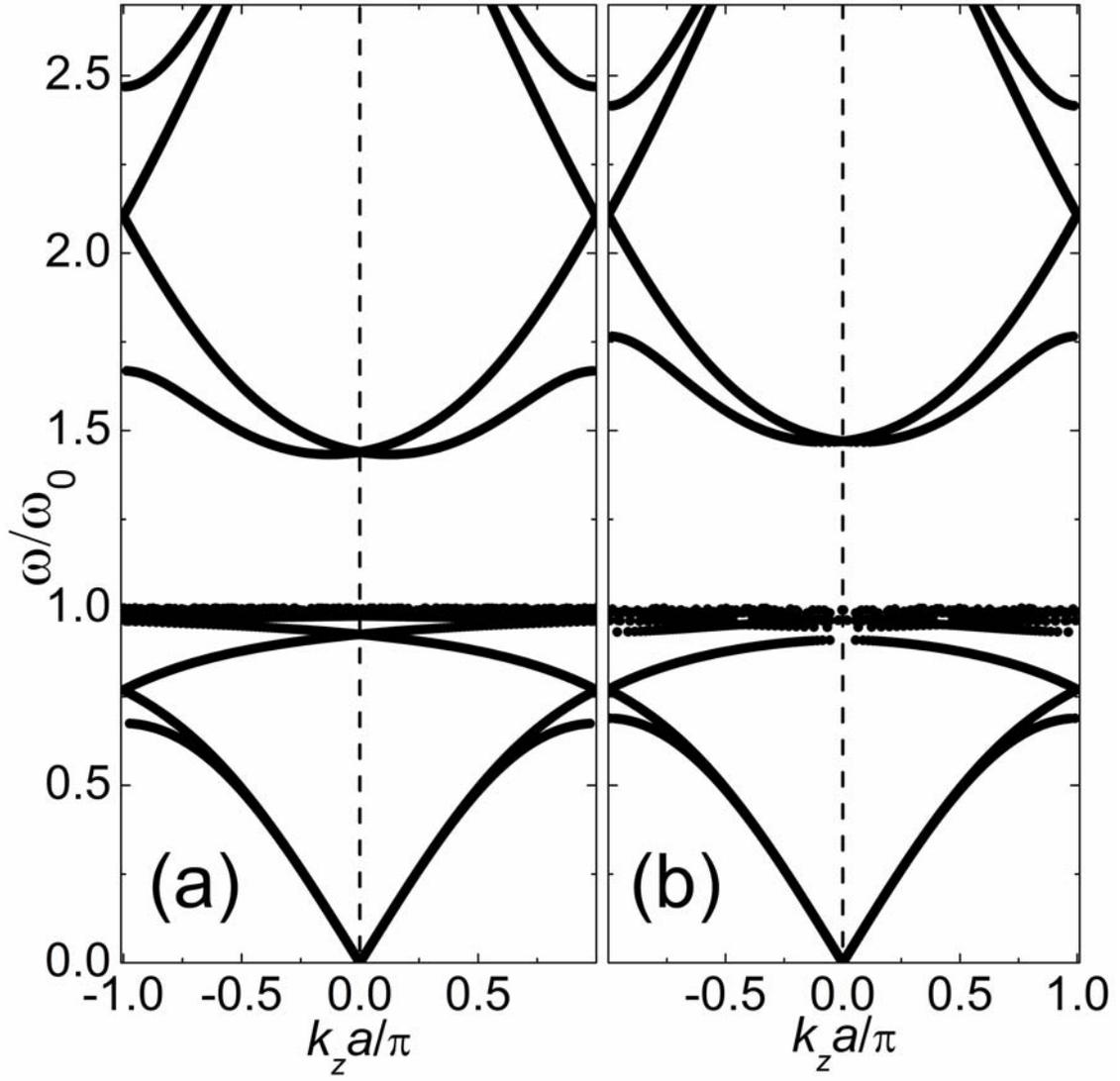

**FIG. 7: Band structures of resonant chiral PCs at** $s = 2.78$ **for (a) continuously twisted resonant chiral PC and (b)** $60°$ **layer-by-layer resonant chiral PC.**

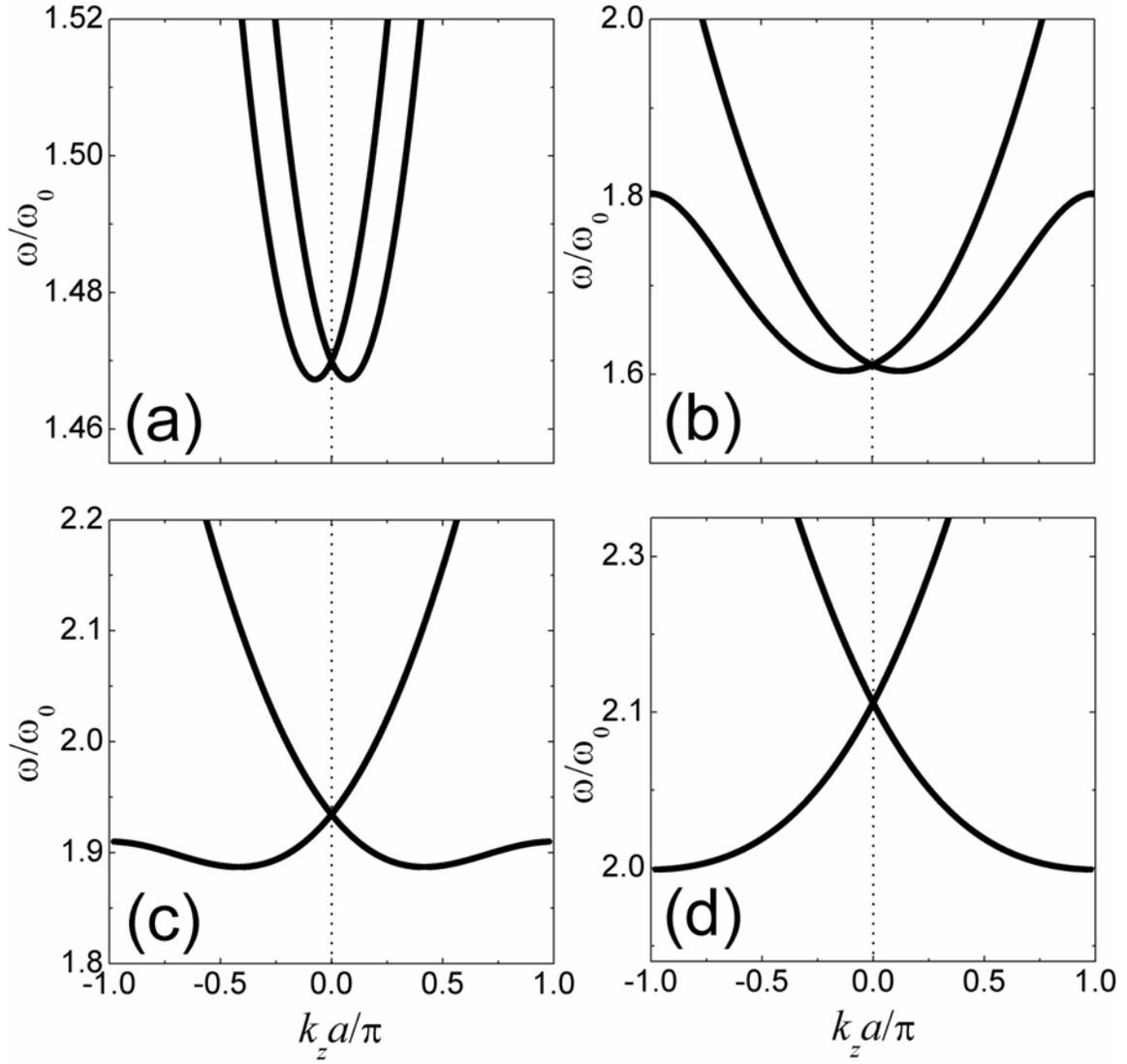

**FIG. 8: The first pass band above the resonance bandgap of the proposed layer-by-layer resonant chiral PC at different resonant strengths. (a)** $s = 2.78$ **(b)** $s = 4$ **(c)** $s = 8$ **(d)** $s = 12$.

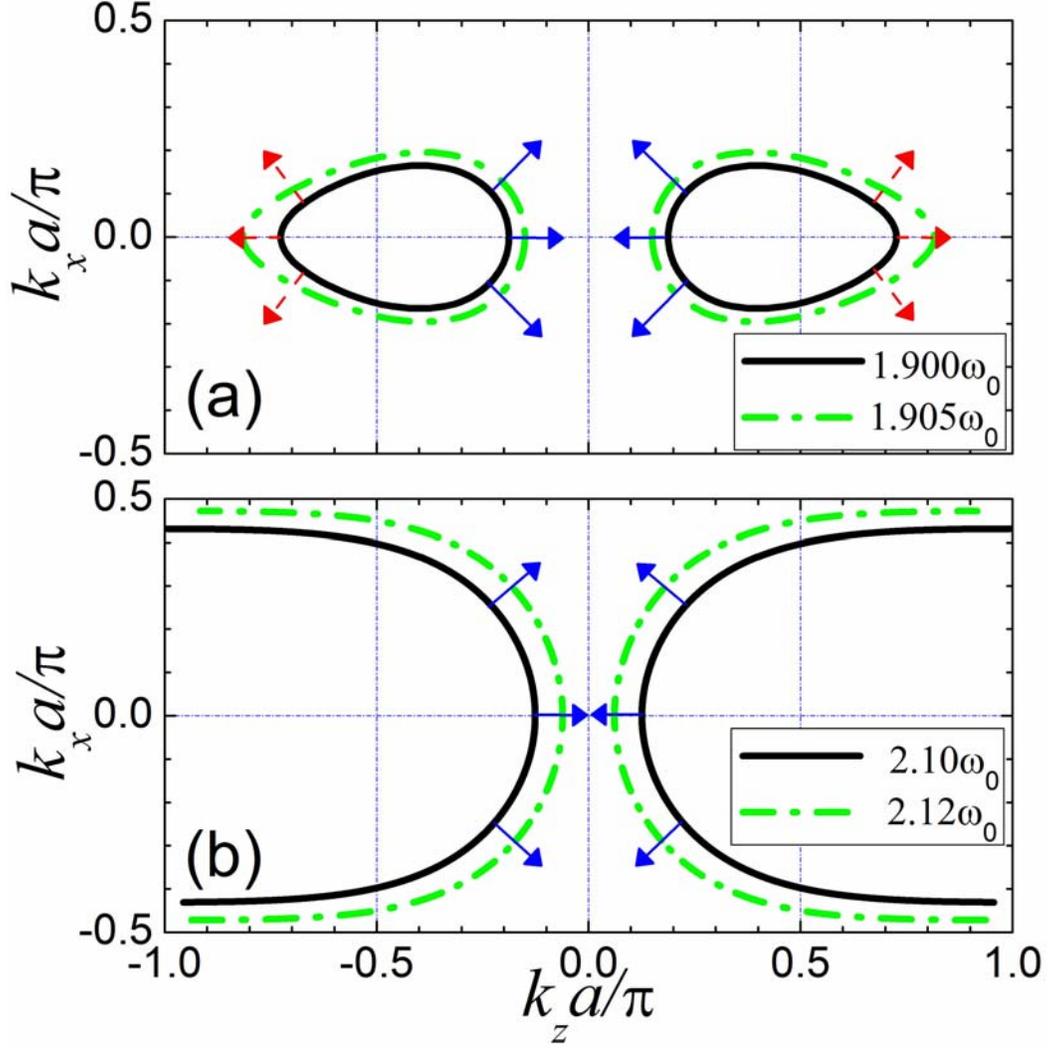

**FIG. 9: (Color online) Iso-frequency contours of the dispersion surface of the proposed layer-by-layer resonant chiral PC in the $k_x - k_z$ plane. Blue (red) arrows show the directions of the group velocity of the left-hand (right-hand) polarized modes. (a) $s = 8$. Black solid lines and green dash-dot lines are the contours at $\omega = 1.900\omega_0$ and $\omega = 1.905\omega_0$, respectively. (a) $s = 12$. Black solid lines and green dash-dot lines are the contours at $\omega = 2.10\omega_0$ and $\omega = 2.12\omega_0$, respectively.**

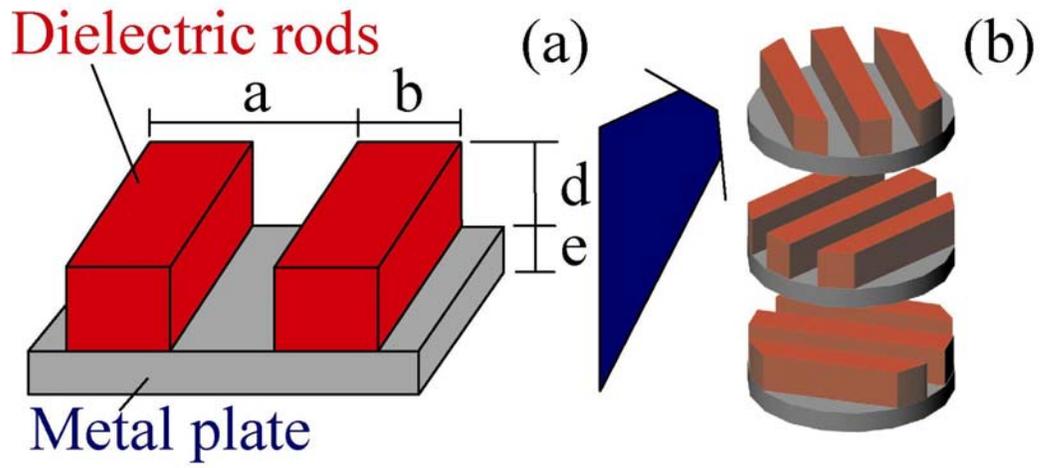

FIG. 10: (Color online) Schematic diagram of the proposed realization of resonant chiral photonic crystal.

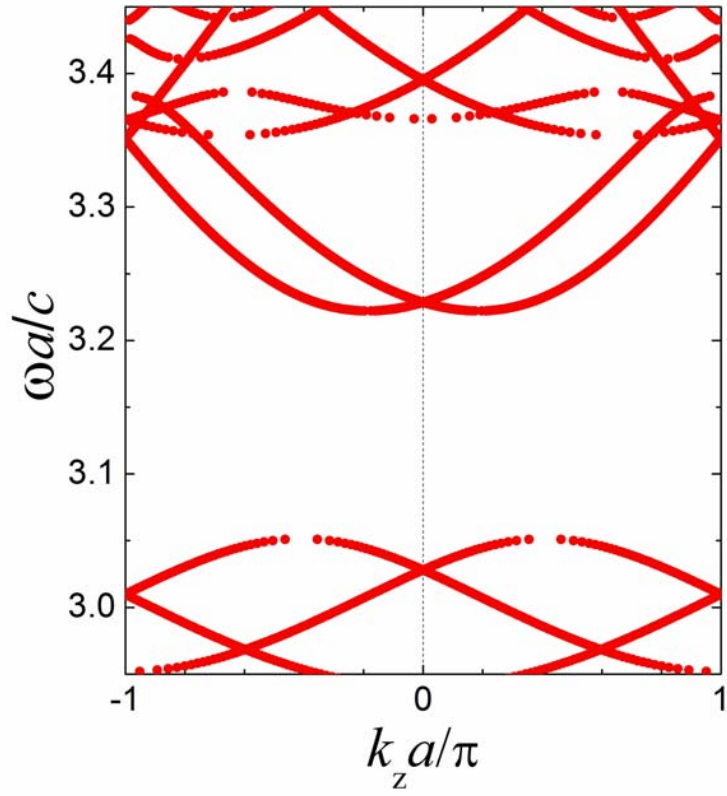

FIG. 11: (Color online) Dispersion relation of the proposed realization as shown in Fig. 10.

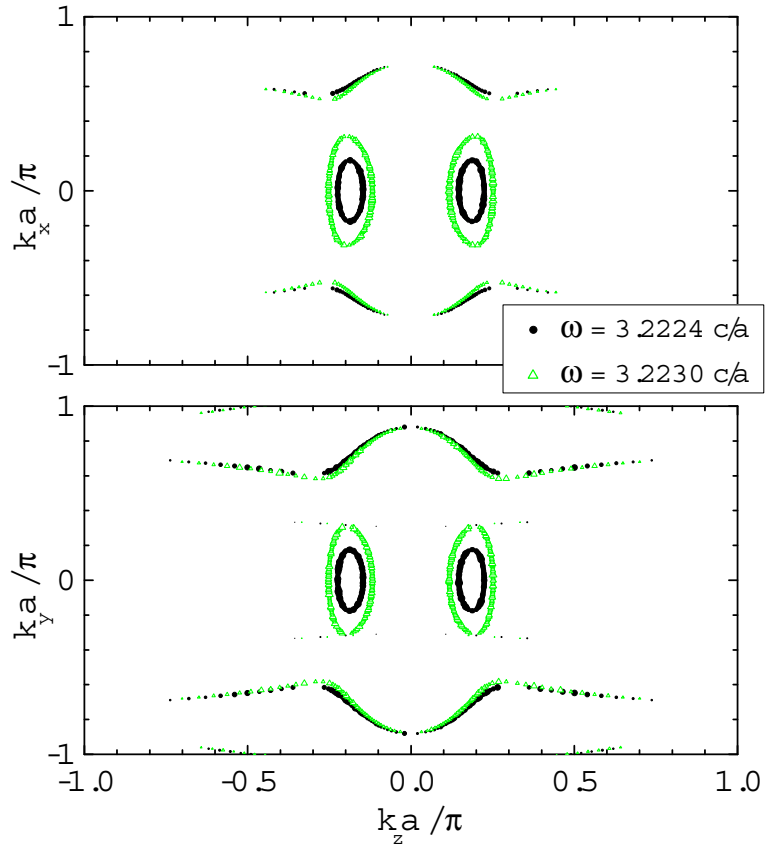

**FIG. 12: (Color online) Iso-frequency contours of the band structures of the proposed realization as shown in Fig. 10.**